\documentclass[epj]{svjour}
\usepackage{graphics}
\usepackage{graphicx}% Include figure files
\usepackage{dcolumn}% Align table columns on decimal point
\usepackage{bm}% bold math
\usepackage[utf8]{inputenc}
\usepackage[T1]{fontenc}
\usepackage{mathptmx}
\usepackage{gensymb}
\newcommand{\eqref}[1]{(\ref{#1})}
\newcommand{\text}[1]{\mbox{\footnotesize #1}}
\begin{document}
\title{Fresnel polarisation of infra-red radiation by elemental bismuth}

\author{Bruno S. C. Alexandre\inst{1} \and Lu\'is C. Martins\inst{2} \and Jaime E. Santos$^\dag$\inst{1} \and Ant\'{o}nio J. Pontes\inst{2} \and Nuno M. R. Peres\inst{1} \inst{3}}
\institute{ Centro de F\'{i}sica, Universidade do Minho, P-4710-057 Braga, Portugal 
 	\and  Instituto de Pol\'imeros e Comp\'ositos, Universidade do Minho, P-4800-058 Guimar\~aes, Portugal
 	\and International Iberian Nanotechnology Laboratory, Avenida Mestre José Veiga s/n, 4715-330 Braga, Portugal
 	\\ \email{$^\dag$jaime.santos@fisica.uminho.pt}
 	}
\date{Received: date / Revised version: date}

\abstract{We revisit the classical problem of electromagnetic wave refraction from a lossless dielectric to a lossy conductor, where both media are considered to be non-magnetic, linear, isotropic and homogeneous. We derive the Fresnel coefficients of the system and the Poynting vectors at the interface, in order to compute the reflectance and transmittance of the system. We use a particular parametrisation of the referred Fresnel coefficients so as to make a connection with the ones obtained for refraction by an interface between two lossless media. This analysis allows the discussion of an actual application, namely the Fresnel polarisation of infra-red radiation by elemental bismuth, based on the concept of pseudo Brewster's angle.
}%end of abstract
\maketitle

%\begin{quotation}
%The ``lead paragraph'' is encapsulated with the \LaTeX\ 
%\verb+quotation+ environment and is formatted as a single paragraph before the first section heading. 
%(The \verb+quotation+ environment reverts to its usual meaning after the first sectioning command.) 
%Note that numbered references are allowed in the lead paragraph.
%%
%The lead paragraph will only be found in an article being prepared for the journal \textit{Chaos}.
%\end{quotation}

\section{Introduction}

Propagation of electromagnetic (EM) waves is a central problem in electromagnetism and optics. The most elementary scattering problem involving electromagnetic waves is that of refraction by a plane interface separating two dielectric media with distinct electrical and/or magnetic properties \cite{griffiths,Jackson,Born-Wolf}. One can study more complex problems by considering that one or both media are dissipative \cite{Dupertuis,Frezza,Vitela,Kim,Canning,Shen,Zang,Oughstun} and/or dispersive \cite{Churchill} and generalize for multiple media transmission \cite{Weber,Lin}. Frezza et al. \cite{Frezza} investigate the transmission of a plane wave between two lossy media and show the absence of total-reflection in these systems, whereas Kim et al. \cite{Kim} considered the transmission of light from a transparent medium onto an absorbing substrate. They specifically derive an analytical solution for the pseudo Brewster's angle, that is, the angle of incidence of the incoming wave for which one of the polarised components of the reflected wave shows a minimum of intensity (unlike the case of two dielectric media for which such a minimum attains a zero value, in which case one speaks of a true Brewster's angle).

Reflection of EM waves by such an angle is thus a method for obtaining polarised light, the so-called Fresnel polarisation method. Dummer et al. \cite{Dummer} designed and constructed a linear polariser consisting of four germanium plates arranged in a chevron geometry that can be used in a frequency band going from infrared to visible radiation ($0.4$ $\mu m$ to $500$ $\mu m$). They claimed that germanium is the appropriate material for such a device given its high and nearly constant refraction index in the range of frequencies considered. 

Other methods of polarisation of IR radiation have been discussed in the literature. A polariser built with stacks of ali\-gned single-walled carbon nanotubes \cite{Ren} is ideal for terahertz applications with a polarization degree of $99.9\%$. Its performance is due to the inherent anisotropic THz absorption properties of aligned SWCNTs (single-walled carbon nanotubes). Huang et al. \cite{Huang} presented a TM/TE wave splitter composed of a gyrotropic slab where light polarization is obtained by transmission instead of reflection. This medium is both anisotropic and non-reciprocal under an applied DC magnetic field and depending on the working frequency band and on the applied magnetic field, total reflection can occur at the boundary of the slab for either TE or TM component of the incident waves. It is also possible to have a total transmission of the polarized wave if the incident angle is selected appropriately and the degree of polarisation increases with the thickness of the slab.

Moreover, the most common way to polarize THz radiation is through wire-grid polarisers \cite{Costley,Kondo,Novak}, which are required to have a narrow and uniform pitch to achieve a high extinction ratio in this frequency region \cite{Yamada}. In order to enhance polarization of this waves, Yamada et al. \cite{Yamada} fabricated a micrometer-pitch Al grating on a Si substrate by photo-lithography. The reason for choosing Al as the wire material is its lower resistivity, that leads to a higher extinction coefficient in the terahertz region. At the Brewster's angle, the transmittance for TM polarization exceeds $95\%$, which is a sign of the high efficiency of this type of polarisers.

In this paper, we revisit the problem of refraction of an electromagnetic plane wave by a boundary separating two non-magnetic media, the first of which is
a lossless dielectric and the second a lossy conductor. We will pay considerable attention to the notion of a pseudo Brewster's angle introduced above. 

Despite the pedagogical character of the paper, our main purpose is to discuss an actual application, namely the Fresnel polarisation of infra-red radiation by elemental bismuth, whose dielectric function shows the appropriate characteristics within the relevant spectral range for such an endeavour \cite{Toudert}. 

We would like to point out that the need to revisit the problem of refraction at an interface between a lossless and a lossy medium is justified by the fact that bismuth behaves as a metal in essentially the whole of the frequency-range studied (see below), behaving as a dielectric in a much narrower sub-range of frequencies where its efficiency as a polariser is the highest, such a narrower sub-range being the one we wish to identify for practical purposes. Moreover, the parametrisation introduced to described the fields in the lossy medium allows us to highlight the differences between refraction by an interface separating two dielectrics and the case that we wish to discuss. This is a distinct approach to the use of a complex refractive index and a complex transmission angle to describe the properties of the lossy medium \cite{Kim,Chen}. 

We start with Maxwell equations in linear, isotropic and homogeneous media in order to obtain the electric and magnetic fields and discuss the boundary conditions
at the interface between the two media. After the  kinematics of the refraction process has been discussed, i.e. the relation between the wave-vectors of the incident and transmitted and reflected waves, we compute the Fresnel coefficients, namely the ratio between the amplitude of the transmitted or the reflected fields with the amplitude of the incident wave. We will also consider the expression for the Poynting vectors of the electromagnetic field in the two media, so as to obtain the reflectance and transmittance at the interface in terms of the said Fresnel coefficients.

This paper is organized as follows: in section \ref{secDR}, we use Maxwell equations in order to obtain the dispersion relations in both media. In section \ref{secRT}, we write the expressions for the electric fields in all regions of space and, using the appropriate boundary conditions, we generalize Snell's law and derive the Fresnel coefficients for our problem. This calculation is performed using a particular parametrisation for the field in the lossy material in terms of real angles. In section \ref{secPR}, we compute the power relations at the interface so as to obtain the reflectance and transmittance of the system. In section \ref{secPF}, we use the results discussed in the previous sections to study Fresnel polarisation on reflection by elemental bismuth within the infra-red frequency range. In section \ref{secConc}, we present our conclusions. Finally, in appendix
\ref{appO}, we apply our results in the context of Drude's classical theory of conduction in metals and discuss the concept of Brewster's angle in doped Si, which is well described by this approximation and also has, like Bi, a large real part of the dielectric function in the relevant frequency range. In appendix \ref{appA}, we express the Poynting's vector in the lossy medium in terms of the magnetic field instead of the electric field, as in the main text and discuss TE and TM like modes in lossy media. In appendix \ref{appB}, we prove two useful identities used in section \ref{secPR}.

\section{Plane waves in lossless and lossy media: dispersion relations}
\label{secDR}
In the following, we shall consider the refraction of a plane EM wave on a plane interface, which
separates a lossless medium with relative permittivity  $\varepsilon_{1r}$ and relative permeability $\mu_{1r}$, and a lossy medium
with relative permittivity  $\varepsilon_{2r}$, relative permeability $\mu_{2r}$, and conductivity $\sigma_2$ (see figure \ref{image0}). The SI units are used throughout
and all the above constants are supposed real, with $\varepsilon_0$ and $\mu_0$ being, respectively, the permittivity and permeability of the vacuum. 
We consider a coordinate system in which the said interface coincides with the $z=0$ plane, so that its normal is oriented along the $zz$ axis.

The homogeneous Maxwell equations, describing the propagation of plane waves are given, for each of these two media, in the frequency domain, by
\begin{eqnarray}
	\nabla\cdot\mathbf{H}&=&0\,,
	\label{eqMax1}
	\\
	\nabla\times\mathbf{E}&=&i\mu_0\mu_r\omega\mathbf{H}\,,
	\label{eqMax2}
	\\
	\nabla\cdot\mathbf{E}&=&0\,,
	\label{eqMax3}
	\\
	\nabla\times\mathbf{H}&=&-i\varepsilon_0\varepsilon_{c}\omega\mathbf{E}\,,
	\label{eqMax4}
\end{eqnarray}
where $\mu_r=\mu_{1,2r}$ depending on the medium, but $\varepsilon_{c}=\varepsilon_{1r}$ in the case of medium 1, whereas
$\varepsilon_c=\varepsilon_{2c}=\varepsilon_{2r}(1+i\tau_2)$, in the case of medium 2, where $\tau_2(\omega)=\frac{\sigma_2}{\omega\varepsilon_0\varepsilon_{2r}}$ is
the loss tangent of the said medium. Given the homogeneous character of the problem, the presence of a conduction current in medium 2 can be included in the description by considering a complex permittivity with the above form \cite{Jackson}. 

We assume for simplicity that the conductivity of medium 2 is real, however, in the case where 
such conductivity is a complex function, i.e. $\sigma_2(\omega)=\sigma_2^{'}(\omega)+i\sigma_2^{''}(\omega)$, one may still apply what is said  below by substituting
\begin{equation}
\varepsilon_{2r}\rightarrow\tilde{\varepsilon}_{2r}(\omega)=\varepsilon_{2r}\left(1-\frac{\sigma_2^{''}(\omega)}{\omega\varepsilon_0\varepsilon_{2r}}\right)\,,
\label{eqepsilonomega}
\end{equation}
 and
\begin{equation}
\tau_{2}(\omega)\rightarrow\tilde{\tau}_{2}(\omega)=\frac{\sigma_2^{'}(\omega)}{\omega\varepsilon_0\varepsilon_{2r}-\sigma_2^{''}(\omega)}\,,
\label{tauomega}
\end{equation}
so there is no loss of generality in supposing $\sigma_2$ to be real. 
Moreover, note that $\text{Re} [\varepsilon_{2c}(\omega)]=\tilde{\varepsilon}_{2r}(\omega)$, $\text{Im} [\varepsilon_{2c}(\omega)]=\frac{\sigma_2^{'}(\omega)}{\omega\varepsilon_0}$, and therefore $\tilde{\tau}_{2}(\omega)=\frac{\text{Im} [\varepsilon_{2c}(\omega)]}{\text{Re} [\varepsilon_{2c}(\omega)]}$, just as before. 

In section \ref{secPF}, we apply the results of our analysis to the element bismuth by using the functional form of the dielectric function discussed by Toudert and co-workers \cite{Toudert}, which is a fit of their experimental data and from which it follows that the conductivity of this material indeed possess a real and imaginary part, dependent on the frequency. 

In appendix \ref{appO}, we discuss the semi-classical Drude approximation, in which the conductivity also possesses frequency dependent real and imaginary parts. Despite its simplicity, it accurately describes the optical properties of selected materials. We apply the results obtained to the analysis of doped Si, which shares some characteristics with Bi in the relevant frequency range. 

In order to obtain the solution of Maxwell's equations, one considers plane-waves with the usual form
\begin{eqnarray}
	\mathbf{H}(\mathbf{r})&=&\mathbf{H}_0\,e^{i\mathbf{k}\cdot\mathbf{r}}\,,
	\label{eqPW1}
	\\
	\mathbf{E}(\mathbf{r})&=&\mathbf{E}_0\,e^{i\mathbf{k}\cdot\mathbf{r}}\,,
	\label{eqPW2}
\end{eqnarray}
and substitutes these expressions in equations \eqref{eqMax1} to \eqref{eqMax4} above, obtaining
\begin{eqnarray}
	\mathbf{k}\cdot\mathbf{H}_0&=&0\,,
	\label{eqMaxPW1}
	\\
	\mathbf{k}\times\mathbf{E}_0&=&\mu_0\mu_r\omega\mathbf{H}_0\,,
	\label{eqMaxPW2}
	\\
	\mathbf{k}\cdot\mathbf{E}_0&=&0\,,
	\label{eqMaxPW3}
	\\
	\mathbf{k}\times\mathbf{H}_0&=&-\varepsilon_0\varepsilon_{2c}\omega\mathbf{E}_0\,.
	\label{eqMaxPW4}
\end{eqnarray}
Equations \eqref{eqMaxPW1} and \eqref{eqMaxPW3} express the transverse character of the fields, whereas equations \eqref{eqMaxPW2} and \eqref{eqMaxPW4} relate 
the amplitudes $\mathbf{H}_0$ and $\mathbf{E}_0$ to each other. Substituting, e.g. the expression for $\mathbf{H}_0$, as given by \eqref{eqMaxPW2}, in \eqref{eqMaxPW4} and using the transverse condition for the electric field, one obtains the dispersion equation for the waves, which is the condition for the existence of non-zero solutions of the homogeneous Maxwell equations,
\begin{equation}
	\mathbf{k}\cdot\mathbf{k}=\frac{\omega^2}{c^2}\mu_{r}\varepsilon_{c}\,,
	\label{eqdisp}
\end{equation}
where $c=\frac{1}{\sqrt{\mu_{0}\varepsilon_0}}$ is the speed of light in the vacuum.

In the case of medium 1, $\mathbf{k}_1$ is a real wave vector and we obtain the dispersion relation 
\begin{equation}
	k_1=\frac{\omega}{c}\sqrt{\mu_{1r}\varepsilon_{1r}}\,,
	\label{eqdispls1}
\end{equation}
where $k_1$ is the modulus of the vector $\mathbf{k}_1$. However, in the case of medium 2, this equation implies that the wave vector, $\mathbf{k}_2=\mathbf{k}^{'}_2+i\mathbf{k}^{''}_2$, has a real and an imaginary part and one obtains two equations from \eqref{eqdisp},
\begin{eqnarray}
	k_2^{'2}-k_2^{''2}=\frac{\omega^2}{c^2}\mu_{2r}\varepsilon_{2r}\,,
	\label{eqdisp1}
	\\
	\mathbf{k}_2^{'}\cdot\mathbf{k}_2^{''}=\frac{\omega^2}{2c^2}\mu_{2r}\varepsilon_{2r}\tau_2(\omega)\,,
	\label{eqdisp2}
\end{eqnarray}
where $k_2^{'}$ and $k_2^{''}$ are, respectively, the modulus of $\mathbf{k}_2^{'}$ and of $\mathbf{k}_2^{''}$.
Note that while for medium 1,  one obtains a linear dispersion relation for plane waves, in the case of medium 2 this is not true 
even in the case of a real constant conductivity, as $\tau_2$ is a function of $\omega$.

The vector $\mathbf{k}_2^{'}$ determines the variation of the phase of the wave, whereas $\mathbf{k}_2^{''}$ determines the decay of its amplitude within the 
lossy medium, due to dissipation. If these two vectors are parallel, the wave is called a {\it uniform damped wave}, i.e. the planes of constant phase coincide with those of constant amplitude, whereas otherwise it is called a {\it non-uniform damped wave}.  

A decaying non-uniform wave is possible even in a  lossless medium, i.e. when $\tau_2=0$. From \eqref{eqdisp2}, one sees that in such a case,
the two vectors  $\mathbf{k}_2^{'}$ and $\mathbf{k}_2^{''}$ have to be perpendicular. This solution describes an evanescent wave in the said lossless medium.

Considering the general case of a non-uniform wave and taking the angle between the vectors $\mathbf{k}_2^{'}$ and $\mathbf{k}_2^{''}$ to be given by $\zeta$, if we substitute $k_2^{''}$ in terms of $k_2^{'}$ in \eqref{eqdisp1} using \eqref{eqdisp2}, we obtain the dispersion equation for $k_2^{'}$
\begin{equation}
	k_2^{'4}-\frac{\omega^2\mu_{2r}\varepsilon_{2r}}{c^2}k_2^{'2}-\frac{\omega^4\mu_{2r}^2\varepsilon_{2r}^2\tau_2^2}{4c^4\cos^2\zeta}=0\,,
	\label{eqdisploss}
\end{equation}
with 
\begin{equation}
k_2^{''}=\frac{\omega^2\mu_{2r}\varepsilon_{2r}\tau_2}{2c^2\cos\zeta k_2^{'}}\,.
\label{eqk''}
\end{equation}
Since equation \eqref{eqdisploss} is a quadratic equation for $k_2^{'2}$, its single positive root can be easily extracted. We obtain for $k_2^{'}$ and
$k_2^{''}$, the relations
\begin{eqnarray}
	k_2^{'}&=&\frac{\omega\sqrt{\mu_{2r}\varepsilon_{2r}}}{\sqrt{2}c}\,\left[\left(1+\frac{\tau_2^2}{\cos^2\zeta}\right)^{1/2}+1\right]^{1/2}\,,
	\label{eqkp}
	\\
	k_2^{''}&=&\frac{\omega\sqrt{\mu_{2r}\varepsilon_{2r}}}{\sqrt{2}c}\,\left[\left(1+\frac{\tau_2^2}{\cos^2\zeta}\right)^{1/2}-1\right]^{1/2}\,.
	\label{eqkpp}
\end{eqnarray}

The limiting case of a uniform wave, $\zeta=0$, is simpler to analyse. In the limit 
$\tau_2\ll 1$, we obtain for the penetration depth $\delta_2=1/k_2^{''}$, the result $\delta_2=\frac{2}{\sigma Z_2}$, where $Z_2=\sqrt{\frac{\mu_0\mu_{2r}}{\varepsilon_0\varepsilon_{2r}}}$ is the (real) impedance of the lossy medium. Conversely if
$\tau_2\gg 1$, $\delta_2=\sqrt{\frac{2}{\sigma \omega \mu_0 \mu_{2r}}}$. 

Since the conductivity is a dimensionful quantity, the quantity which determines whether a medium can be considered as a conductor or as a dielectric is the dimensionless loss tangent $\tau$. If $\tau\gg 1$, the medium can be considered a good conductor at the particular frequency we are exciting it, whereas the opposite limit will characterise a dielectric. 

As an example, if the frequency of the radiation  $\nu=3\times 10^{9}\,\text{Hz}$,
one has, for a medium with conductivity $\sigma=10\,\text{Sm}^{-1}$ at that frequency, $\tau \approx 10^3$, i.e. one is in the high $\tau$ limit. In that limit, $\delta\approx 3\,\text{mm}$, applying the above formula, much smaller than the wavelength of the radiation in vacuum. Moreover, in such a limit, the penetration depth is only weakly dependent on the value of the conductivity. An increase of the conductivity by e.g.,  a factor of 25, implies a mere decrease in the penetration depth by a factor of 5. 

In the case where the conductivity is a frequency dependent complex function, we saw above that these results can be carried through if $\epsilon_{2r}$ is substituted by the real part of the dielectric function, and the loss tangent by the ratio of its imaginary and the real parts. Note, however, that if in a particular region of the spectrum, $\varepsilon_{2}^{'}(\omega)<0$, as in the Drude approximation below a certain frequency, see appendix \ref{appO}, the analysis carried above can be taken without any changes, except for equations \eqref{eqkp} and \eqref{eqkpp}. It is easy to check that if we substitute $\varepsilon_{2r}$ by $-|\varepsilon_{2}^{'}(\omega)|$ in equation \eqref{eqdisploss}, we have to interchange $k_2^{'}$ and $k_2^{''}$ in the solutions \eqref{eqkp} and \eqref{eqkpp}. Below, we will use instead an alternative expression for the components of the complex wave-vector where the said substitution can be performed without any caveats.  

\section{Reflection and transmission at the interface between a lossless and lossy medium}
\label{secRT}
We now consider the refraction of a plane wave, incident from the lossless medium, at an angle $\theta_i$ with the normal to the interface. This will give rise to a reflected wave into the lossless medium, at an angle $\theta_r$ with the normal and a transmitted wave into the lossy medium, at an angle $\theta_t$. 

We consider, without loss of generality, that the wave-vectors of the incident and reflect wave are given by 
$\mathbf{k}_{i}=q_i\hat{\mathbf{x}}+u_i\hat{\mathbf{z}}$ and
$\mathbf{k}_{r}=q_r\hat{\mathbf{x}}+u_r\hat{\mathbf{z}}$, i.e. we take the plane of incidence to be the $xz$ plane. 
Both these vectors obey the dispersion relation
\eqref{eqdispls1}. Note that $q_i=k_{i}\sin\theta_i$, $u_i=k_{i}\cos\theta_i$, $q_r=k_{r}\sin\theta_r$ and $u_r=k_{r}\cos\theta_r$, by definition of the 
angles of incidence and reflection.

One writes for the electric field of the incident and reflect wave, the expressions
\begin{eqnarray}
\mathbf{E}_i(\mathbf{r})&=&\mathbf{E}_{i0}e^{i\mathbf{k}_i\cdot\mathbf{r}}=\left[\,E_{ip}(\cos\theta_i\hat{\mathbf{x}}-\sin\theta_i\hat{\mathbf{z}})+E_{is}\hat{\mathbf{y}}\,\right]\nonumber
	\\
	&&\mbox{}\times e^{i(q_ix+u_iz)}\,,
	\label{eqEi}
	\\
\mathbf{E}_r(\mathbf{r})&=&\mathbf{E}_{r0}e^{i\mathbf{k}_r\cdot\mathbf{r}}=\left[\,-E_{rp}(\cos\theta_r\hat{\mathbf{x}}-\sin\theta_r\hat{\mathbf{z}})+E_{rs}\hat{\mathbf{y}}\,\right]\nonumber
	\\
	&&\mbox{}\times e^{i(q_rx+u_rz)},
	\label{eqEr}
\end{eqnarray}
where the subscripts $p$ and $s$ stand, respectively, for the components of the electric field in the plane of incidence and perpendicular to it. By construction
and given the expressions for $q_i$, $u_i$, $q_r$ and $u_r$, both these two vectors obey the transversality condition, equation \eqref{eqMax3}. The choice of signal for the amplitude $E_{rp}$ is such that the polarisation axes of all three waves coincide in the limit of normal incidence, i.e. when $\theta_i=0$, see below.

In the case of the transmitted wave, the lossy character of the medium determines that its wave-vector $\mathbf{k}_{t}=\mathbf{k}_{t}^{'}+i\mathbf{k}_{t}^{''}$ will be composed of a real and imaginary part,
with $\mathbf{k}_{t}^{'}=q_t^{'}\hat{\mathbf{x}}+u_t^{'}\hat{\mathbf{z}}$ and $\mathbf{k}_{t}^{''}=q_t^{''}\hat{\mathbf{x}}+u_t^{''}\hat{\mathbf{z}}$.
We have $q_t^{'}=k_{t}^{'}\sin\theta_t$ and $u_t^{'}=k_{t}^{'}\cos\theta_t$. The two vectors, $\mathbf{k}_{t}^{'}$ and $\mathbf{k}_{t}^{''}$, satisfy the dispersion relations, equations \eqref{eqdisp1} and \eqref{eqdisp2}.

The electric field of the transmitted wave can be written as
\begin{eqnarray}
	\mathbf{E}_t(\mathbf{r})&=&\mathbf{E}_{t0}\,e^{i\mathbf{k}_t\cdot\mathbf{r}}=\left[E_{tp}(\cos\phi_t\hat{\mathbf{x}}-e^{-i\alpha_t}\sin\phi_t\hat{\mathbf{z}})+E_{ts}\hat{\mathbf{y}}\right]\,\nonumber \\ 
	&&\mbox{}\times e^{i(q_t^{'}x+
		u_t^{'}z)}\times e^{-(q_t^{''}x+
		u_t^{''}z)}\,,
	\label{eqEt}
\end{eqnarray}
where we stress that $\alpha_t$ and $\phi_t$ are real parameters, which have to be adequately chosen so as to enforce the transversality condition for $\mathbf{E}_t(\mathbf{r})$, equation \eqref{eqMax3}, given that
the wave vector $\mathbf{k}_{t}$ is complex.  Note, in particular, that one has for the unit complex vector $\hat{\mathbf{v}}_t=\cos\phi_t\hat{\mathbf{x}}-e^{-i\alpha_t}\sin\phi_t\hat{\mathbf{z}}$, $\hat{\mathbf{v}}_t\cdot\hat{\mathbf{v}}_t^*=1$, $\hat{\mathbf{v}}_t\times\hat{\mathbf{v}}_t^*\neq\mathbf{0}$ and $\hat{\mathbf{v}}_t\cdot\hat{\mathbf{v}}_t\neq 0$, where the $^*$ indicates the operation of complex conjugation, which implies that the polarization of the transmitted wave in the plane of incidence is in general {\it elliptical}. The use of the present parametrisation is in contrast with the traditional use of a complex refractive index and a complex transmission angle, as noted above. 

\begin{figure}
	\centering
	%\begin{tabular}{|c|c|c|}
	%	\hline 
	\includegraphics[scale=0.6]{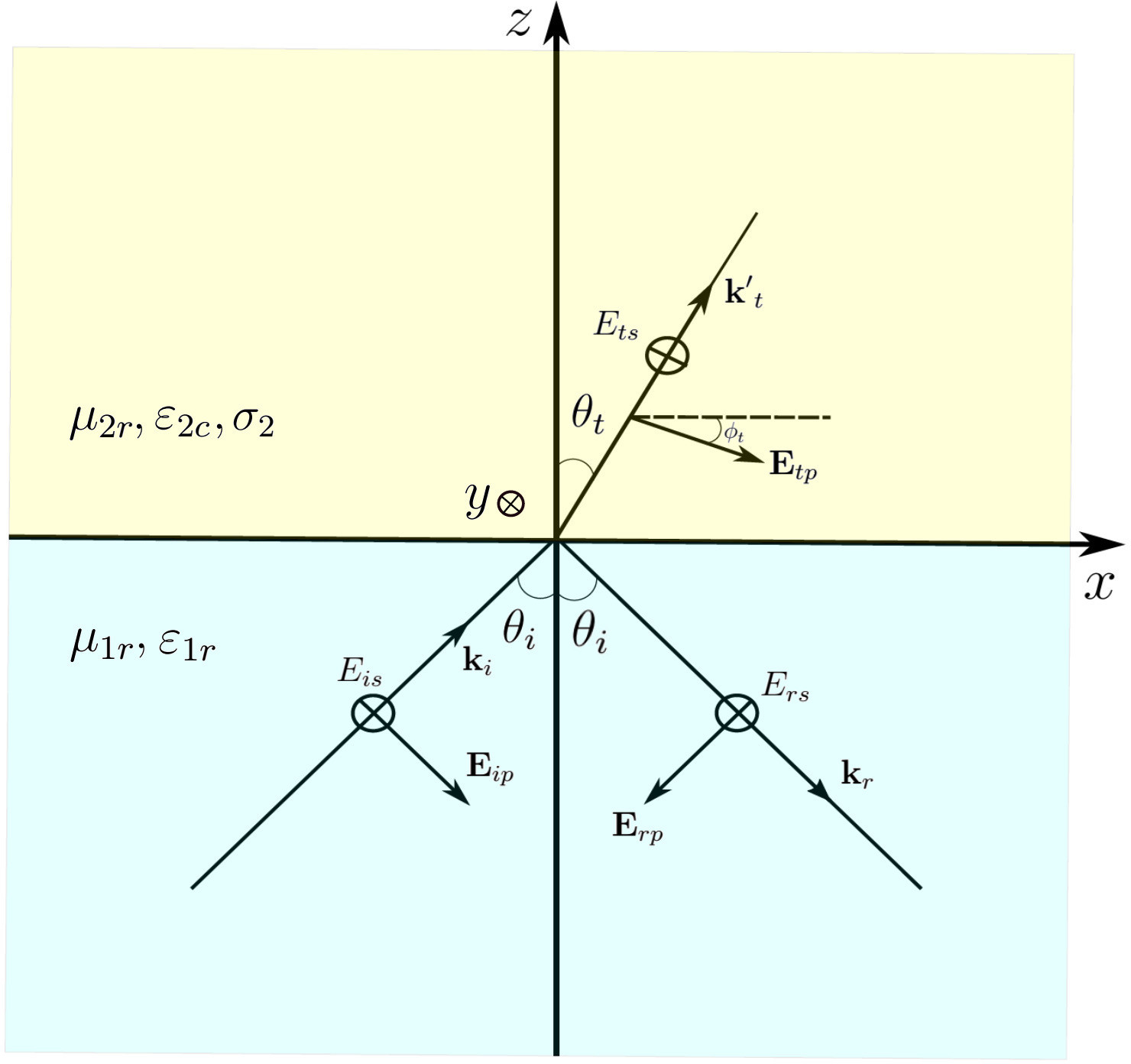} 
	%\tabularnewline
	%\hline 
	%\end{tabular}
	\caption{Refraction by an interface separating a lossless (below) from a lossy medium (above), where the different parameters describing the problem were introduced in the main text. The s-polarisation component of the electric field is along the $y$-axis, into the plane of the figure.}
	\label{image0}
\end{figure}

The different quantities introduced above are represented graphically in figure \ref{image0}, where we show the geometry of the refraction process.

The boundary conditions at the interface that determine the relations between the fields in media 1 and 2 can be obtained from the integral form of
the Maxwell equations and are given by
\begin{eqnarray}
	\hat{\mathbf{z}}\cdot(\mu_{2r}\mathbf{H}_{2}-\mu_{1r}\mathbf{H}_{1})_{z=0}&=&0\,
	\label{eqbc1}
	\\
	\hat{\mathbf{z}}\times(\mathbf{E}_{2}-\mathbf{E}_{1})_{z=0}&=&\mathbf{0}\,,
	\label{eqbc2}
	\\
	\hat{\mathbf{z}}\cdot(\varepsilon_{2c}\mathbf{E}_{2}-\varepsilon_{1r}\mathbf{E}_{1})_{z=0}&=&0\,,
	\label{eqbc3}
	\\
	\hat{\mathbf{z}}\times(\mathbf{H}_{2}-\mathbf{H}_{1})_{z=0}&=&\mathbf{0}\,,
	\label{eqbc4}
\end{eqnarray}
where $\mathbf{E}_{1}(\mathbf{r})=\mathbf{E}_i(\mathbf{r})+\mathbf{E}_r(\mathbf{r})$ is the field in medium 1, $\mathbf{E}_{2}(\mathbf{r})=\mathbf{E}_t(\mathbf{r})$ is the field in medium 2, and where the magnetic field in media 1 and 2 is related to the electric field by equation \eqref{eqMax2}. Since the electric and magnetic fields are not independent, not all the equations \eqref{eqbc1} to \eqref{eqbc4} are independent and we only need the
vector equations \eqref{eqbc2} and \eqref{eqbc4} to determine the solution of the problem. These are four equations determining four unknowns, namely the amplitudes
$E_{rp}$, $E_{rs}$, $E_{tp}$ and $E_{ts}$ ($E_{ip}$ and $E_{is}$ are supposed to be given). Do also note that
if we were considering refraction by a slab of lossy material, the field in medium 2 would also include a reflected component and the third medium would be the 
one where only a transmitted component would exist.

Substituting equations \eqref{eqEi}, \eqref{eqEr} and \eqref{eqEt} in equation \eqref{eqbc2}, we immediately conclude that given that such an equation has to hold for arbitrary values of $x$, $q_t^{'}=q_r=q_i$ and, moreover, that $q_t^{''}=0$. Hence, from this latest result, we conclude that only in the case of normal incidence, $\theta_i=0$, will the wave in medium 2 be uniform. We also conclude that $\theta_r=\pi-\theta_i$, in other words the angle of reflection is equal to the angle of incidence, and that 
\begin{equation}
	k_{t}^{'}\sin\theta_t=\frac{\omega}{c}\,\sqrt{\mu_{1r}\varepsilon_{1r}}\,\sin\theta_i\,,
	\label{eqSnellG}
\end{equation}
which is the generalisation of Snell's law, but where $k_{t}^{'}$ is still to be determined. Note that since $\mathbf{k}_{t}^{''}$ is oriented along the $z$ axis, $\zeta=\theta_t$, thus we can obtain $k_{t}^{'}$ from equation \eqref{eqkp} simply by performing the said substitution. However, this will gives us an equation for $\sin\theta_i$ in terms of the $\sin\theta_t$ and $\cos\theta_t$. Given that it is $\theta_i$ which is normally given (measurable), it is preferable to follow a different route, see also the remark at the end of this section. We have that 
\begin{equation}
	k_{t}^{'2}=q_t^{'2}+u_t^{'2}=\frac{\omega^2}{c^2}\mu_{1r}\varepsilon_{1r}\sin^2\theta_i+u_t^{'2}\,,
	\label{eqdispt2}
\end{equation}
from equation \eqref{eqSnellG}. Substituting this result in equation \eqref{eqdisp1}, together with equation \eqref{eqdisp2}, we obtain
\begin{eqnarray}
	u_t^{'2}-u_t^{''2}=\frac{\omega^2}{c^2}(\mu_{2r}\varepsilon_{2r}-\mu_{1r}\varepsilon_{1r}\sin^2\theta_i)\,,
	\label{eqdisprf1}
	\\
	u_t^{'}u_t^{''}=\frac{\omega^2}{2c^2}\mu_{2r}\varepsilon_{2r}\tau_2(\omega)\,.
	\label{eqdisprf2}
\end{eqnarray}
Substituting $u_t^{''}$ in terms of $u_t^{'}$, as given by equation \eqref{eqdisprf2}, in equation \eqref{eqdisprf1} and solving the resulting quadratic equation for $u_t^{'2}$, we get for $u_t^{'}$ and $u_t^{''}$ the result
\begin{eqnarray}
	u_t^{'}&=&\frac{\omega}{\sqrt{2}c}\,\bigg[\,\sqrt{(\mu_{2r}\varepsilon_{2r}-\mu_{1r}\varepsilon_{1r}\sin^2\theta_i)^2+\mu_{2r}^2\varepsilon_{2r}^2\tau_2^2(\omega)}\nonumber\\ 
	&&\mbox{}+(\mu_{2r}\varepsilon_{2r}-\mu_{1r}\varepsilon_{1r}\sin^2\theta_i)\bigg]^{1/2}\,,
	\label{equp}
\end{eqnarray}

\begin{eqnarray}
	u_t^{''}&=&\frac{\omega}{\sqrt{2}c}\,\bigg[\sqrt{(\mu_{2r}\varepsilon_{2r}-\mu_{1r}\varepsilon_{1r}\sin^2\theta_i)^2+\mu_{2r}^2\varepsilon_{2r}^2\tau_2^2(\omega)}\nonumber \\ 
	&&\mbox{}-(\mu_{2r}\varepsilon_{2r}-\mu_{1r}\varepsilon_{1r}\sin^2\theta_i)\bigg]^{1/2}\,.
	\label{equpp}
\end{eqnarray}

We can in turn substitute equation \eqref{equp} in equation \eqref{eqdispt2}, obtaining
\begin{eqnarray}
	k_{t}^{'}&=&\frac{\omega}{\sqrt{2}c}\left[\,\sqrt{(\mu_{2r}\varepsilon_{2r}-\mu_{1r}\varepsilon_{1r}\sin^2\theta_i)^2+\mu_{2r}^2\varepsilon_{2r}^2\tau_2^2(\omega)}\right.\nonumber\\
	&&\mbox{}\left.+(\mu_{2r}\varepsilon_{2r}+\mu_{1r}\varepsilon_{1r}\sin^2\theta_i)\,\right]^{1/2}
	\label{eqk1tF}
\end{eqnarray}
Using such result, we finally obtain from \eqref{eqSnellG} for $\sin\theta_t$ , the expression
\begin{eqnarray}
	\sin\theta_t&=&\sqrt{2\mu_{1r}\varepsilon_{1r}}\sin\theta_i\,\big\{\,
	\left[\,(\mu_{2r}\varepsilon_{2r}-\mu_{1r}\varepsilon_{1r}\sin^2\theta_i)^2+\right.\\
	&&\mbox{}\left.\mu_{2r}^2\varepsilon_{2r}^2\tau_2^2(\omega)\,\right]^{1/2}+\mu_{2r}\varepsilon_{2r}+\mu_{1r}\varepsilon_{1r}\sin^2\theta_i\,\big\}^{-1/2}\,,\nonumber
	\label{eqSnellG2}
	\end{eqnarray}
in terms of $\sin\theta_i$. It reduces to the known result for Snell's law when $\tau_2=0$. Note that this result can be loosely interpreted as if medium 2 had an index of refraction that depends on the direction of incidence of the incoming wave.

From the condition of transversality $\nabla\cdot\mathbf{E}_t(\mathbf{r})=0$, we obtain a complex equation relating $\alpha_t$ and $\phi_t$,
\begin{equation}
	q_t^{'}e^{i\alpha_t}=(u_t^{'}+iu_t^{''})\tan\phi_t\,.
	\label{eqtransv1}
\end{equation}
Separating the real and imaginary part of this equation and expressing $q_t^{'}$ and $u_t^{'}$ in terms of $k_t^{'}$ and $\theta_t$, we obtain

\begin{eqnarray}
	\tan\phi_t&=&\tan\theta_t\cos\alpha_t\,,
	\label{eqphit}
	\\
	\tan\alpha_t&=&\frac{u_t^{''}}{u_t^{'}}=\frac{\omega^2\mu_{2r}\varepsilon_{2r}\tau_2(\omega)}{2c^2u_t^{'2}}\\
	&=&\mu_{2r}\varepsilon_{2r}\tau_2(\omega)\bigg\{
	[(\mu_{2r}\varepsilon_{2r}-\mu_{1r}\varepsilon_{1r}\sin^2\theta_i)^2\nonumber\\
	&&\mbox{}+\mu_{2r}^2\varepsilon_{2r}^2\tau_2^2(\omega)]^{1/2}+\mu_{2r}\varepsilon_{2r}-\mu_{1r}\varepsilon_{1r}\sin^2\theta_i\bigg\}^{-1/2}\,,\nonumber
	\label{eqalphat}
\end{eqnarray}
where we used equations \eqref{eqdisprf2} and \eqref{equp}. If $\tau_2=0$, we obtain $\alpha_t=0$ and $\phi_t=\theta_t$, as one would expect. Note that from 
equation \eqref{eqphit} it follows that $\phi_t\leq \theta_t$. 

It should be observed from equation \eqref{eqSnellG2} that, regardless of the ratio of $\mu_{1r}\varepsilon_{1r}$ to $\mu_{2r}\varepsilon_{2r}$, one always
has $\sin\theta_t<1$. This is in contrast to the case of two dielectrics for which $\mu_{1r}\varepsilon_{1r}>\mu_{2r}\varepsilon_{2r}$. In that case, for
angles of incidence $\theta_i$ such that $\sin\theta_i\geq \sqrt{\frac{\mu_{2r}\varepsilon_{2r}}{\mu_{1r}\varepsilon_{1r}}}$, one observes the phenomenon of {\it total reflection}. Note, in fact, that at large $\tau_2$, $\theta_t\rightarrow 0$, with the penetration depth $\delta _2$ going to zero as discussed above. 

Substituting $\mathbf{E}_{1}(\mathbf{r})$ and $\mathbf{E}_{2}(\mathbf{r})$ in \eqref{eqbc2} and $\mathbf{H}_{1}(\mathbf{r})=\frac{-i}{\mu_0\mu_{1r}\omega}
\nabla\times\mathbf{E}_{1}(\mathbf{r})$ and $\mathbf{H}_{2}(\mathbf{r})=\frac{-i}{\mu_0\mu_{2r}\omega}\nabla\times\mathbf{E}_{2}(\mathbf{r})$ in
\eqref{eqbc4}, one obtains two decoupled systems of equations involving $E_{tp}$ and $E_{rp}$ in terms of $E_{ip}$ and
$E_{ts}$ and $E_{rs}$ in terms of $E_{is}$, which when solved, yield for the transmission and reflection coefficients $t_p=E_{tp}/E_{ip}$, $r_p=E_{rp}/E_{ip}$,
$t_s=E_{ts}/E_{is}$ and $r_s=E_{rs}/E_{is}$, the expressions
\begin{eqnarray}
	t_p&=&\frac{2\zeta_p\cos\theta_i}{\zeta_p\cos\phi_t+Z_1\cos\theta_i}\,,
	\label{eqtp}
	\\
	r_p&=&\frac{\zeta_p\cos\phi_t-Z_1\cos\theta_i}{\zeta_p\cos\phi_t+Z_1\cos\theta_i}\,,
	\label{eqrp}
	\\
	t_s&=&\frac{2\zeta_s\cos\theta_i}{\zeta_s\cos\theta_i+Z_1\cos\phi_t}\,,
	\label{eqts}
	\\
	r_s&=&\frac{\zeta_s\cos\theta_i-Z_1\cos\phi_t}{\zeta_s\cos\theta_i+Z_1\cos\phi_t}\,,
	\label{eqrs}
\end{eqnarray}
where $Z_1=\sqrt{\frac{\mu_0\mu_{1r}}{\varepsilon_0\varepsilon_{1r}}}$ is the impedance of medium 1 and
\begin{eqnarray}
	\zeta_p&=&\frac{u_t^{'}+iu_t^{''}}{\omega\varepsilon_{2c}\varepsilon_0\cos\phi_t},
	\label{eqzp0}
	\\
	\zeta_s&=&\frac{\mu_0\mu_{2r}\omega\cos\phi_t}{u_t^{'}+iu_t^{''}}\,.
	\label{eqzs0}
\end{eqnarray}
Introducing the complex quantity, $x_t=\frac{c (u_t^{'}+iu_t^{''})}{\omega \sqrt{\mu_{2r}\varepsilon_{2c}}}$,  we can also express equations \eqref{eqzp0} and \eqref{eqzs0} as
\begin{eqnarray}
	\zeta_p&=&Z_{2c}\,\frac{x_t}{\cos\phi_t}\,,
	\label{eqzp1}
	\\
	\zeta_s&=&Z_{2c}\,\frac{\cos\phi_t}{x_t}\,,
	\label{eqzs1}
\end{eqnarray}
where $Z_{2c}=\sqrt{\frac{\mu_0\mu_{2r}}{\varepsilon_0\varepsilon_{2c}}}$ is the complex impedance of medium 2. Note that the product, 
$\zeta_p\zeta_s=Z_{2c}^2$, as in the case of two lossless media. If $\theta_i\neq 0$, one can simplify equations \eqref{eqzp0} and \eqref{eqzs0} using equation \eqref{eqtransv1}, to read
\begin{eqnarray}
	\zeta_p&=&Z_{2c}\frac{\sqrt{\mu_{1r}\varepsilon_{1r}}\sin\theta_i}{\sqrt{\mu_{2r}\varepsilon_{2c}}\sin\phi_t}e^{i\alpha_t},
	\label{eqzp}
	\\
	\zeta_s&=&Z_{2c}\frac{\sqrt{\mu_{2r}\varepsilon_{2c}}\sin\phi_t}{\sqrt{\mu_{1r}\varepsilon_{1r}}\sin\theta_i}\,e^{-i\alpha_t}\,,
	\label{eqzs}
\end{eqnarray}
where the relevant parameters entering these formulas were all given above, even if in an implicit form. 
These quantities act as generalised impedances for the lossy medium. It should be noted that they depend on the direction of incidence of the incoming wave and are not to be identified with the complex impedance of the medium, except in the case of normal incidence (see below).

In the limit $\tau_2=0$, with $\phi_t=\theta_t$, $\alpha_t=0$,
we simply obtain, using Snell's law, $\zeta_p=\zeta_s=Z_2$, and the formulas above, equations \eqref{eqtp} to \eqref{eqrs}, 
reduce to the Fresnel formulas for refraction of waves at the interface between two lossless media. 

For the case of normal incidence, $\theta_i=\theta_t=\phi_t=0$, we have, substituting equations \eqref{equp} and \eqref{equpp} in equations \eqref{eqzp1} and \eqref{eqzs1}, in light of the definition of $x_t$, that $x_t=1$ and thus, $\zeta_p=\zeta_s=Z_{2c}$. Again, when $\tau_2=0$, we obtain $\zeta_p=\zeta_s=Z_2$. Moreover, it can be seen from equation \eqref{eqEt} that, as $\phi_t=0$, the angle $\alpha_t$ completely drops out of the analysis. Also, note that since in this case $\theta_r=\pi$, one sees from equations \eqref{eqEi}, \eqref{eqEr}  and \eqref{eqEt} that the polarization axes of the incident, reflected and transmitted waves all coincide. 

Note that in all the manipulations performed in this section,  we could have substituted $\varepsilon_{2r}\rightarrow \varepsilon^{'}_2(\omega)$ and $\tau_2(\omega)\rightarrow \tilde{\tau}_2(\omega)$ with these quantities either both positive or negative (if $\varepsilon^{'}_2(\omega)<0$). This represents another advantage of this procedure to the direct use of equation \eqref{eqkp} in equation \eqref{eqSnellG}.

Also, note from equations \eqref{eqrp} and \eqref{eqrs} that, because $\zeta_p$ and $\zeta_s$ are complex quantities, there isn't generally a Brewster's angle in the refraction between a dielectric and a conductive media, such that either $r_p$ or $r_s$ are zero, depending on the relative value of the impedances, as happens in the refraction at the boundary between two dielectric media. There is still, nevertheless an angle for which the amplitude of $r_p$ reaches a non-zero minimum and one speaks in this case of a {\it pseudo Brewster's} angle. 

\section{Power relations at the interface}
\label{secPR}
In order to consider the power relations at the interface, we need to compute the expressions for the averaged power of the field over a period of the incoming wave, at each side of the said interface. We have for the Poynting vector of the EM field, the expression
\begin{equation}
	\mathbf{\cal P}_1(\mathbf{r})=\left\langle\mathbf{\cal E}_1(\mathbf{r},t)\times\mathbf{\cal H}_1(\mathbf{r},t)\right\rangle\,,
	\label{eqPoyn1}
\end{equation}
for medium 1, where $\mathbf{\cal E}_1(\mathbf{r},t)=\text{Re}\left(\mathbf{E}_1(\mathbf{r},t)e^{-i\omega t}\right)$ and 
$\mathbf{\cal H}_1(\mathbf{r},t)=\text{Re}\left(\mathbf{H}_1(\mathbf{r},t)e^{-i\omega t}\right)$ and
\begin{equation}
	\mathbf{\cal P}_2(\mathbf{r})=\left\langle\mathbf{\cal E}_2(\mathbf{r},t)\times\mathbf{\cal H}_2(\mathbf{r},t)\right\rangle\,,
	\label{eqPoyn2}
\end{equation}
for medium 2, where $\mathbf{\cal E}_2(\mathbf{r},t)=\text{Re}\left(\mathbf{E}_2(\mathbf{r},t)e^{-i\omega t}\right)$ and 
$\mathbf{\cal H}_2(\mathbf{r},t)=\text{Re}\left(\mathbf{H}_2(\mathbf{r},t)e^{-i\omega t}\right)$ (in determining physical quantities, such as an energy flux, we use the real part of the expressions we have determined so far, to compute the physical electric and magnetic fields). 
Substituting the known expressions
for these fields as given by equations \eqref{eqEi}, \eqref{eqEr} and \eqref{eqEt}, we obtain after averaging over a period 
of the incoming wave, for the Poynting vector in medium 1 \cite{Chen}

\begin{eqnarray}
	\mathbf{\cal P}_1(\mathbf{r})&=&\frac{1}{2Z_1}\,\left[|\mathbf{E}_{i0}|^2\hat{\mathbf{k}}_i+|\mathbf{E}_{r0}|^2\hat{\mathbf{k}}_r\right]\nonumber\\  
	&&\mbox{}+\frac{1}{2\omega\mu_0\mu_{1r}}\bigg[\,\text{Re}\left(\mathbf{E}_{i0}\cdot\mathbf{E}_{r0}^{*}\right)(\mathbf{k}_i+\mathbf{k}_r)\nonumber\\  
	&&\mbox{}+\text{Re}\left(\mathbf{E}_{i0}\cdot(\mathbf{k}_i-\mathbf{k}_r)\mathbf{E}_{r0}^{*}\right)\nonumber\\  
	&&\mbox{}-\text{Re}\left(\mathbf{E}_{r0}\cdot(\mathbf{k}_i-\mathbf{k}_r)\mathbf{E}_{i0}^{*}\right)\bigg]\,,
	\label{eqP1}
\end{eqnarray}
where $\hat{\mathbf{k}}_i$ and $\hat{\mathbf{k}}_r$ are the unit vectors in the direction of incidence of the incoming and reflected wave, respectively. Note that in this particular case, the Poynting vector is independent of the position, indicating that energy is not absorbed by medium 1. 

In medium 2, we obtain
\begin{eqnarray}
	\mathbf{\cal P}_2(\mathbf{r})&=&\frac{e^{-2u_t^{''}z}}{2\omega\mu_0\mu_{2r}}\left[\mathbf{k}_t^{'}|\mathbf{E}_{t0}|^2-iu_t^{''}\hat{\mathbf{z}}\times\left(\mathbf{E}_{t0}\times\mathbf{E}_{t0}^{*}\right)\right]\,,
	\label{eqP2}
\end{eqnarray}
where the decay of the Poynting vector with the distance from the interface indicates that energy is being absorbed in medium 2
(in appendix \ref{appA}, this quantity is expressed in terms of the magnetic field). Using the formulas \eqref{eqtp} to
\eqref{eqrs}, one can show from equations \eqref{eqP1} and \eqref{eqP2} that

\begin{eqnarray}
&&\hat{\mathbf{z}}\cdot\left[\mathbf{\cal P}_2(\mathbf{r})-\mathbf{\cal P}_1(\mathbf{r})\right]_{z=0}\nonumber\\
&=&\frac{2\cos^2\theta_i\cos\phi_t|\zeta_p|^2}{|Z_1\cos\theta_i+\zeta_p\cos\phi_t|^2}\left[\,\frac{k_t^{'}\cos\theta_t}{\omega\mu_0\mu_{2r}\cos\phi_t}-\text{Re}\left(\zeta_p^{-1}\right)\,\right]|E_{ip}|^2 \nonumber\\
&&\mbox{}+\frac{2\cos^2\theta_i\cos\phi_t|\zeta_s|^2}{|Z_1\cos\phi_t+\zeta_p\cos\theta_i|^2}\,\left[\,\frac{k_t^{'}\cos\theta_t}{\omega\mu_0\mu_{2r}\cos\phi_t}-\text{Re}\left(\zeta_s^{-1}\right)\,\right]|E_{is}|^2\nonumber\\
&=& 0\,.
\label{EQP3}
\end{eqnarray}
The identities for the real parts of the inverse permittivities that were used in equation \eqref{EQP3} are shown in appendix \ref{appB}. 

We can define the coefficients 
\begin{equation}
{\cal R}=\frac{|\mathbf{E}_{r0}|^2}{|\mathbf{E}_{i0}|^2}
\end{equation}
and 
\begin{equation}
{\cal T}=\frac{\mu_{1r}\tan\theta_i|\mathbf{E}_{t0}|^2}{\mu_{2r}\tan{\theta}_t|\mathbf{E}_{i0}|^2},
\end{equation}
which represent the relative amount of transmitted and reflected energy at the interface. From equation \eqref{EQP3}, one has ${\cal T}+{\cal R}=1$, with
\begin{eqnarray}
	{\cal R}&=&|r_p|^2\cos^2\xi+|r_s|^2\sin^2\xi\,,
	\label{eqR}\\
	{\cal T}&=&\frac{\mu_{1r}\tan\theta_i}{\mu_{2r}\tan{\theta}_t}\left(|t_p|^2\cos^2\xi+|t_s|^2\sin^2\xi\right)\,,
	\label{eqT}
\end{eqnarray} 
with $\cos\xi=\frac{|E_{ip}|}{|\mathbf{E}_{i0}|}$ and $\sin\xi=\frac{|E_{is}|}{|\mathbf{E}_{i0}|}$ and where the reflection and transmission coefficients are given by equations \eqref{eqtp} to \eqref{eqrs}.
In the case of normal incidence, we simply have ${\cal R}=|r|^2$ and ${\cal T}=Z_1\,\text{Re}(Z_{2c}^{-1})\,|t|^2$, where $r_p=r_s=r=\frac{Z_{2c}-Z_1}{Z_{2c}+Z_1}$ and $t_p=t_s=\frac{2Z_{2c}}{Z_{2c}+Z_1}$.

\section{Fresnel Polarisation in Elemental Bismuth}
\label{secPF}
\begin{figure}
	\centering
	\begin{tabular}{c}
		%	\hline 
		\includegraphics[scale=0.35]{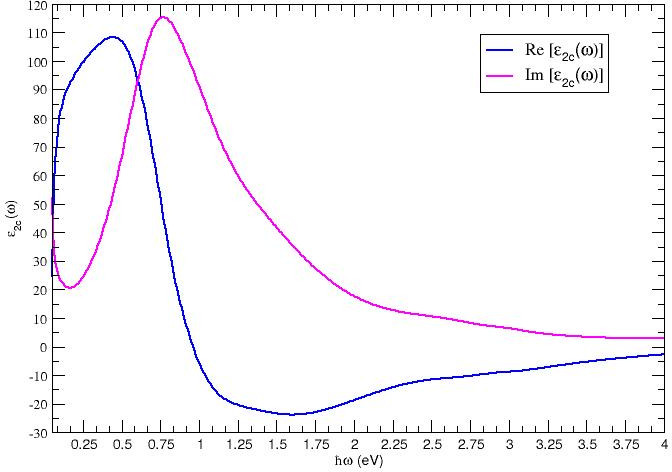} 
		%\tabularnewline
		%	\hline 
	\end{tabular}
	\caption{Dielectric function of elemental bismuth as a function of photon energy. The blue line represents the real part and the magenta line represents the imaginary part of this function.}
	\label{epsilon}
\end{figure}

We will now use the results discussed above to consider Fresnel polarisation, i.e. the appearance of a true rather than a pseudo Brewster's angle, in a real material, more specifically elemental bismuth, in the infra-red frequency range. The functional form of the dielectric function of Bi which we will use is that discussed by Toudert and co-workers \cite{Toudert}, which is a fit of their experimental data using a superposition of Drude and Lorentz-like terms. Such a function is shown in figure \ref{epsilon}, within the range of validity of the fit. 

\begin{figure}
	\centering
	\begin{tabular}{c}
		%	\hline 
		\includegraphics[scale=0.35]{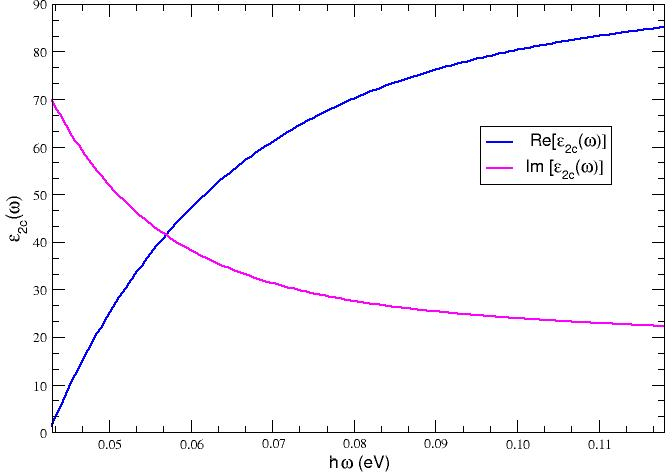} 
		%\tabularnewline
		%	\hline 
	\end{tabular}
	\caption{Dielectric function of elemental bismuth as a function of photon energy. The blue line represents the real part and the magenta line represents the imaginary part of this complex function.}
	\label{epsilon-fig}
\end{figure}

\begin{figure}
	\centering
	\begin{tabular}{c}
		%	\hline 
		\includegraphics[scale=0.35]{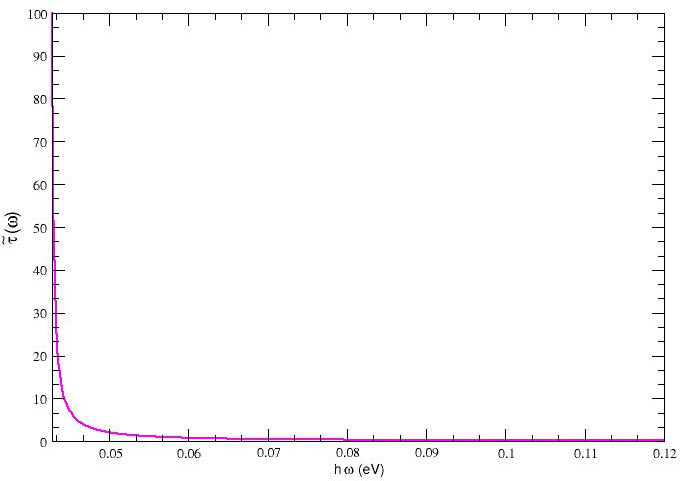}
		%\tabularnewline
		%	\hline 
	\end{tabular}
	\caption{Loss tangent of bismuth in the same range of frequencies as in figure \ref{epsilon-fig}.}
	\label{tau}
\end{figure}

Since we wish to consider the far-infrared region of the electromagnetic spectrum, we also plot both the dielectric function and the loss tangent of this metal within a more restricted range of frequencies, as seen in figure \ref{epsilon-fig} and figure \ref{tau}. One can observe that since the real part of the dielectric function is much larger than its imaginary part for energies in the range $70-120$ meV, Bi is characterised by a small value of the loss tangent and thus behaves essentially as a dielectric, which displays a true Brewster's angle, as discussed above. 

On the other hand, for energies below $50$ meV, the imaginary part is the larger one and we expect Bi to behave as a metal, leading to the appearance of a pseudo Brewster's angle, in which the reflected $p$ component of the radiation impinging on a vacuum-Bi interface will show a minimum value that is larger than zero.

\begin{figure}
	\centering
	\begin{tabular}{ccc}
		%		\hline 
		\includegraphics[scale=0.35]{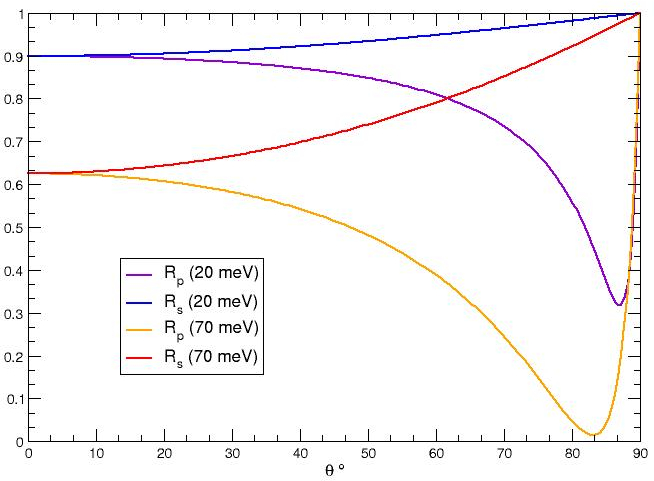} 
		%\tabularnewline
		%		\hline 
	\end{tabular}
	\caption{Values of the reflectance for the s and p components ($R_s$ and $R_p$) at an interface vacuum/elemental bismuth, as a function of the incidence angle ($\theta$), for two different values of the energy of the incoming radiation.}
	\label{Rps}
\end{figure}

In figure \ref{Rps}, we show the graphical illustration of the above discussion, where for energies above 70 meV the reflected p component, $R_p$, becomes approximately zero, whereas for an energy of 20 meV, it simply reaches a finite minimum. 

Moreover, one can see from figure \ref{Rps} that the value of the Brewster's angle is rather large, which can be understood from the fact that the impedance of the medium from where the radiation is incident (vacuum) is significantly larger that of bismuth. It should be further noted that the present method is characterised by a high theoretical efficiency, as the value of $R_s$ is close to one at the Brewster's angle, where $R_p$ is very nearly zero. The properties here discussed are quite similar to those observed for Germanium \cite{Dummer}, for the same reasons. 

The discussion undertaken in this section is the main result of this paper. 

The direct study of polarisation by reflection in Bi thin-films, using light with 632.8 nm, has been reported in the literature \cite{Espinosa2012}, the authors having found a value for the pseudo-Brewster's angle of around 66\degree, whereas our formulas yield a value of around 78\degree. Nevertheless, the authors of \cite{Espinosa2012} have cautioned that due to surface effects, the refraction index of Bi is reduced in the films with respect to the bulk material, from which one indeed expects a reduction of the value of the pseudo-Brewster's angle.

\section{Conclusions}
\label{secConc}

In this paper, we studied the problem of light refraction at an interface separating a lossless dielectric from a lossy conductive medium. We derived the Fresnel coefficients for this system using a particular parametrisation that sheds light on the similarities and differences of this problem with that of refraction between two lossless media. We also explored the concept of pseudo Brewster's angle, the angle for which the $p$-polarised component of reflectance has a minimum when radiation is impinging on the lossy medium. Using this concept, we discussed Fresnel polarisation, which occurs in a region of frequencies where the loss tangent of the dissipative material is small and hence where it behaves as a dielectric, so that a real Brewster's angle emerges.

We considered the Fresnel polarisation by a real material, elemental bismuth, and using an interpolation of its dielectric function obtained in reference \cite{Toudert}, from their experimental data, were able to compute the reflected $p$ and $s$ components of radiation as a function of the incident angle in the infra-red frequency range. Within the energy range $70-120$ meV, Bi behaves as a dielectric since the real part of the dielectric function is much larger than its imaginary part, which leads to a small loss tangent. Thus, this material displays a real Brewster's angle and acts as a highly efficient IR polariser ($R_p\approx 0$, while $R_s\approx 1$), as a result of the small value of its impedance relative to that of the vacuum. 

\section{Authors contributions}
B. Alexandre performed the analysis of the polarisation by Fresnel reflection of bismuth in section \ref{secPF} and derived the results in Appendix \ref{appA}. J. Santos introduced the parametrisation of the electric field used in section \ref{secRT} and performed the analysis presented in appendix \ref{appO}. N. Peres suggested the search of an efficient polarising medium in the far-infrared. All the authors contributed to the remainder of the manuscript and participated in the bibliographic search as well as in its redaction and the checking of results. All the authors approved the final manus\-cript.

\begin{acknowledgement}

We acknowledge helpful discussions with M. Vasilevskiy, P. Alpuim, J. Caridad and B. Figueiredo. 
The authors thank the European Structural and Investment Funds in the FEDER component,
through the Operational Programme for Competitiveness and Internationalization (COMPETE 2020)
[under the Project GNESIS – Graphenest’s New Engineered
System and its Implementation Solutions; Funding Reference: POCI-01-0247-FE\-DER-033566], European Regional
Development Fund. This work was also supported by the Portuguese Foundation for Science and Technology (FCT) in the framework of the Strategic Funding UID/FIS/04650/\-2019.

\end{acknowledgement}

\appendix
\section{The Drude approximation}
\label{appO}
The Drude-Sommerfeld approximation treats charge carriers in conductors as quasi-free fermions which interact with static impurities through collisions, with a mean-time between collisions equal to $\tau_D$. This gives rise to the following form of the AC conductivity
\begin{equation}
\sigma(\omega)=\frac{\sigma_0}{1-i\omega\tau_D}\,,
\label{eqDr}
\end{equation}
with $\sigma_0=\frac{ne^2\tau_D}{m^*}$ being the DC conductivity of the metal, where $n$ is the charge carrier density, $e$ the electron charge and $m^*$ is the carrier's effective mass. This approximation describes well the behaviour of selected three dimensional conductors or doped semi-conductors and is also valid for a two-dimensional semi-metal such as charged graphene within a given frequency range, if we write $m^*=\frac{\hbar k_F}{v_F}$ where $v_F$ is the Fermi velocity of graphene and $k_F=\sqrt{\pi n}$ is the modulus of the Fermi wave-vector, see \cite{Goncalves16} and references therein. It can be derived by elementary means and justified using the semi-classical Boltzmann equation \cite{AshMerm}. 

Substituting the functional form given by equation \eqref{eqDr} in  the equation for the complex dielectric permittivity of a 3d material, $\varepsilon(\omega)=\varepsilon_\infty+\frac{i\sigma(\omega)}{\varepsilon_0\omega}$, at finite frequency \cite{Jackson}, where $\varepsilon_\infty$ is the dielectric constant of the material at infinite frequency, we obtain, introducing the plasma frequency of the material,
$\omega_p=\sqrt{\frac{ne^2}{m^*\varepsilon_0}}$, and the collision rate, $\gamma=1/\tau_D$, for the real and imaginary parts of this function
\begin{eqnarray}
\varepsilon^{'}(\omega)&=&\varepsilon_\infty-\frac{\omega_p^2}{\omega^2+\gamma^2}\,,
\label{eqeDr1}
\\
\varepsilon^{''}(\omega)&=&\frac{\gamma\omega_p^2}{\omega(\omega^2+\gamma^2)}\,.
\label{eqeDr2}
\end{eqnarray}
These expressions can now be used directly in the formulas obtained in section \ref{secRT} to compute the transmission and reflections coefficients of a plane electromagnetic wave undergoing refraction at the interface between a dielectric and a Drude metal. 

Note that from equation \eqref{eqeDr1}, one has that $\varepsilon^{'}(\omega)<0$, if $\omega<\sqrt{\frac{\omega_p^2}{\varepsilon_\infty}-\gamma^2}$ (with 
$\omega_p>\sqrt{\varepsilon_\infty}\gamma$). As pointed out in section \ref{secDR}, one can still use all the results discussed above in this case. 

A three dimensional material which is well described by the Drude approximation in a wide range of frequencies is moderately doped Si \cite{Exter1990}. For p-doped Si with $n=1.4\times 10^{22} \mbox{m}^{-3}$, we have $\omega_p=7.24\, \mbox{meV}$ (THz region), with $\gamma=0.863\omega_p$. The index of refraction for undoped Si is $n_r=3.415$ in the whole  frequency range, from which we obtain $\varepsilon_\infty=n_r^2=11.7$.  With these values, the real  part of the dielectric constant is positive throughout the whole frequency range. 

\begin{figure}
	\centering
	\begin{tabular}{c}
		%	\hline 
		\includegraphics[scale=0.35]{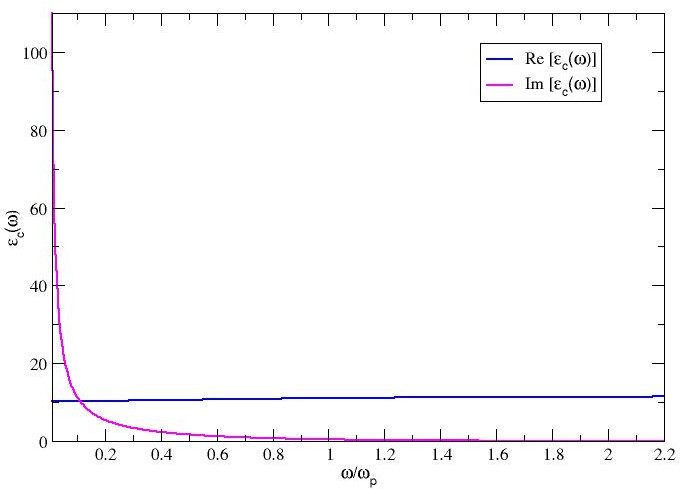} 
		%\tabularnewline
		%	\hline 
	\end{tabular}
	\caption{Dielectric function of elemental Si as a function of frequency (in units of $\omega_p=7.24 \,\mbox{meV}$). The blue line represents the real part and the magenta line represents the imaginary part of this complex function.}
	\label{epsilon-figSi}
\end{figure}

\begin{figure}
	\centering
	\begin{tabular}{c}
		%	\hline 
		\includegraphics[scale=0.35]{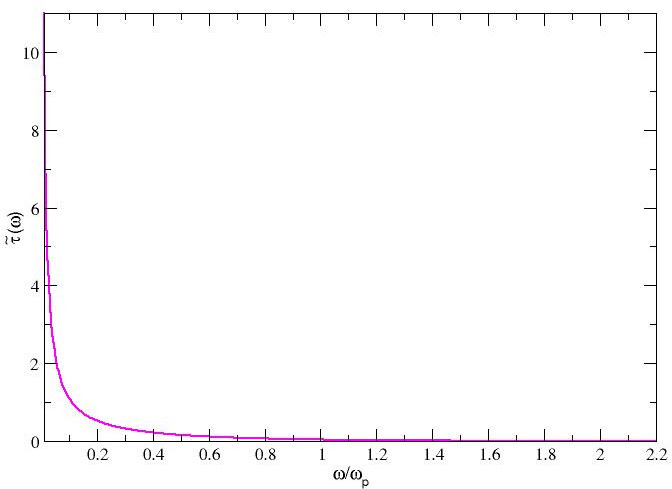}
		%\tabularnewline
		%	\hline 
	\end{tabular}
	\caption{Loss tangent of Si in the same range of energies as in figure \ref{epsilon-figSi}.}
	\label{tauSi}
\end{figure}

\begin{figure}
	\centering
	\begin{tabular}{ccc}
		%		\hline 
		\includegraphics[scale=0.35]{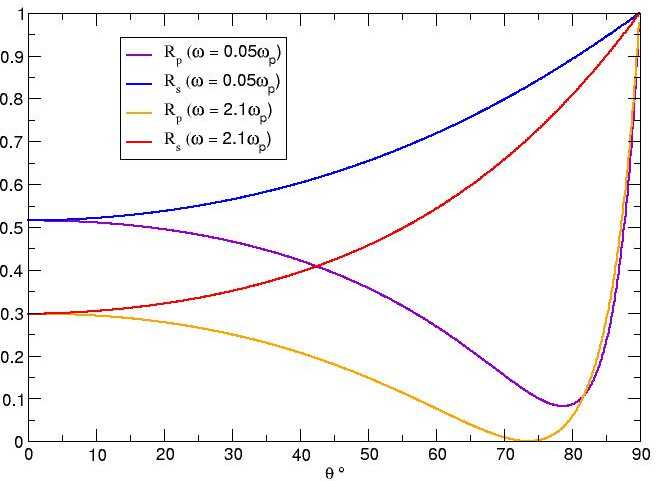} 
		%\tabularnewline
		%		\hline 
	\end{tabular}
	\caption{Values of the reflectance for the s and p components ($R_s$ and $R_p$) at an interface vacuum/p-doped Si, as a function of the incidence angle ($\theta$), for two different values of the frequency of the incoming radiation.}
	\label{RpsSi}
\end{figure}

We plot the real and imaginary parts of the dielectric function (fig. \ref{epsilon-figSi}), as well as the loss tangent (fig. \ref{tauSi}), and the reflectance components (fig. \ref{RpsSi}) at two distinct frequencies, $\omega=0.05\omega_p$ and $\omega=2.1\omega_p$. It is clear from figure \ref{tauSi}, that the material behaves as a metal for frequencies below $0.1\omega_p$, since $\tilde{\tau}(\omega)>1$ in that region, with the material displaying a pseudo Brewster's angle as shown in figure \ref{RpsSi}. Above such frequency, the material behaves as a dielectric and displays a Brewster's angle, as also shown in figure \ref{RpsSi}. 

This behaviour is qualitatively similar to that found for Bi in section \ref{secPF}. Note that given the value of the plasma frequency, it occurs in an energy range that is one order of magnitude below the corresponding one for Bi. Therefore, one could use Si to polarise incoming THz radiation using the same concept discussed for Bi. However, given the value of $R_s$ at the Brewster's angle for $\omega=2.1\omega_p$, the efficiency of such a polariser would be smaller than that of a polariser using Bi, a fact which can be easily understood, given that the relative impedance of Si with respect to the vacuum is larger than that of Bi (the real part of the dielectric function is smaller in the corresponding range of frequencies). 

\section{Poynting's vector in lossy media}
\label{appA}
In this appendix, we express the Poynting's vector in medium 2 in terms of the magnetic field, instead of the electric field, as given by
equation \eqref{eqP2} and discuss its form in the particular case of linearly polarised waves in lossy media. 
We start with the expression for the time-averaged Poynting's vector 
\begin{eqnarray}
\mathbf{\cal P}_2(\mathbf{r})&=&\frac{1}{2}\text{Re}(\mathbf{E}_t(\mathbf{r})\times \mathbf{H}^*_t(\mathbf{r}))\,,
\label{eqP2H}
\end{eqnarray}
where we substitute equation \eqref{eqMaxPW4} in medium 2
\begin{eqnarray}
\mathbf{E}_t(\mathbf{r})&=&-\frac{1}{\epsilon_0\epsilon_c\omega}\mathbf{k}_t\times\mathbf{H}_t(\mathbf{r})\,,
\label{eqA1}
\end{eqnarray}
so as to obtain

\begin{equation}
\mathbf{\cal P}_2(\mathbf{r})=\frac{1}{2\varepsilon_0\omega}\text{Re}\left(\frac{|\mathbf{H}_{t0}|^2\,\mathbf{k}_t-(\mathbf{k}_t\cdot\mathbf{H}_{t0}^*)\mathbf{H}_{t0}}{\varepsilon_{2c}}\right)e^{-2\mathbf{k}_t^{''}\cdot\mathbf{r}}.
\label{eqA2}
\end{equation}

This can be rewritten as

\begin{eqnarray}
\mathbf{\cal P}_2(\mathbf{r})=\frac{1}{2\varepsilon_0\omega|\varepsilon_{2c}|^2}\bigg\{
|\mathbf{H}_{t0}|^2(\varepsilon_{2c}^{'}\mathbf{k}_t^{'}+\varepsilon_{2c}^{''}\mathbf{k}_t^{''}) \nonumber
\end{eqnarray}
\begin{equation}
-\frac{1}{2}[\,\varepsilon_{2c}^*(\mathbf{k}_t\cdot\mathbf{H}_{t0}^*)\mathbf{H}_{t0}+\varepsilon_{2c}(\mathbf{k}_t^*\cdot\mathbf{H}_{t0})\mathbf{H}_{t0}^*]
\bigg\}\,e^{-2\mathbf{k}_t^{''}\cdot\mathbf{r}}\,.
\label{eqA3}
\end{equation}

It is possible to simplify the second and third terms of the above equation using the identity

\begin{eqnarray}
\varepsilon_{2c}^*(\mathbf{k}_t\cdot\mathbf{H}_{t0}^*)\mathbf{H}_{t0}+\varepsilon_{2c}(\mathbf{k}_t^*\cdot\mathbf{H}_{t0})\mathbf{H}_{t0}^*=\varepsilon_{2c}^{'}[\,(\mathbf{k}_t\cdot\mathbf{H}_{t0}^*)\mathbf{H}_{t0}\nonumber\\ 
+(\mathbf{k}_t^*\cdot\mathbf{H}_{t0})\mathbf{H}_{t0}^*\,]
+i\varepsilon_{2c}^{''}[\,(\mathbf{k}_t^*\cdot\mathbf{H}_{t0})\mathbf{H}_{t0}^*-(\mathbf{k}_t\cdot\mathbf{H}_{t0}^*)\mathbf{H}_{t0}\,]\,.
\label{eqA4}
\end{eqnarray}

Using equation \eqref{eqMaxPW1} and its complex conjugate, we can write the terms in the rhs of the last equation as

\begin{equation}
(\mathbf{k}_t\cdot\mathbf{H}_{t0}^*)\mathbf{H}_{t0}+(\mathbf{k}_t^*\cdot\mathbf{H}_{t0})\mathbf{H}_{t0}^*=(\mathbf{k}_t-\mathbf{k}_t^*)\times(\mathbf{H}_{t0}\times\mathbf{H}_{t0}^*),
\label{eqA5}
\end{equation}
\begin{equation}
(\mathbf{k}_t\cdot\mathbf{H}_{t0}^*)\mathbf{H}_{t0}-(\mathbf{k}_t^*\cdot\mathbf{H}_{t0})\mathbf{H}_{t0}^*=(\mathbf{k}_t+\mathbf{k}_t^*)\times(\mathbf{H}_{t0}\times\mathbf{H}_{t0}^*),
\label{eqA6}
\end{equation}
arriving finally at the expression for Poynting's vector
\begin{eqnarray}
\mathbf{\cal P}_2(\mathbf{r})&=&\frac{1}{2\varepsilon_0\omega|\varepsilon_{2c}|^2}\bigg[|\mathbf{H}_{t0}|^2(\varepsilon_{2c}^{'}\mathbf{k}_t^{'}+\varepsilon_{2c}^{''}u_t^{''}\mathbf{z})\nonumber\\
&&\mbox{}+i(\varepsilon_{2c}^{''}\mathbf{k}_t^{'}-\varepsilon_{2c}^{'}u_t^{''}\mathbf{z})\times(\mathbf{H}_{t0}\times\mathbf{H}_{t0}^*)\bigg]e^{-2u_t^{''}z}\,.
\label{eqP2HH}
\end{eqnarray}

One can show \cite{Chen} that simultaneous linear polarisation of both the electric and the magnetic field in a lossy media cannot occur.
However, in the case where the electric field is linearly polarised, $\mathbf{E}_{t0}^{*}=e^{i\delta}\mathbf{E}_{t0}$, where $\delta$ is a real phase, 
and therefore $\mathbf{E}_{t0}\times
\mathbf{E}_{t0}^{*}=\mathbf{0}$, and hence the second term of the rhs of equation \eqref{eqP2} is zero. Since it follows from equation \eqref{eqMaxPW3} and its complex conjugate that $\mathbf{k}_t^{'}\cdot\mathbf{E}_{t0}=0$ and $\mathbf{k}_t^{''}\cdot\mathbf{E}_{t0}=0$, when the electric field is linearly polarised, one concludes that in this case the Poynting vector is perpendicular 
to $\mathbf{E}_{t0}$. The same conclusion follows for $\mathbf{H}_{t0}$ from \eqref{eqP2HH}, if the magnetic field is linearly polarised. These modes can thus be interpreted as the TE and TM modes in a lossy medium. Note, however that, as pointed out in the main text, fields resulting from the process of refraction are in general elliptically polarised. Linear polarisation would occur if either the electric or the magnetic field of the incident wave were perpendicular to the plane of incidence, giving rise in the material to the said TE or TM modes referred above \cite{Vitela}.
\section{Identities used to show equation \eqref{EQP3}}
\label{appB}
Here we derive the identities for the inverse permittivities required to show equation \eqref{EQP3}. Starting from equation \eqref{eqzp0}, we write, using the dispersion equation \eqref{eqdisp} in medium 2 in a complex form
\begin{eqnarray}
\zeta_p^{-1}&=&\frac{\omega\varepsilon_0\varepsilon_{2c}\cos\phi_t}{u_t^{'}+iu_t^{''}}=\frac{\cos\phi_t}{\mu_0\mu_{2r}\omega}\cdot\frac{q_t^{'2}+(u_t^{'}+iu_t^{''})^2}{u_t^{'}+iu_t^{''}}\nonumber\\
&=&\frac{\cos\phi_t}{\mu_0\mu_{2r}\omega}\cdot\left[\,q_t^{'}\tan\phi_t \,e^{-i\alpha_t}+(u_t^{'}+iu_t^{''})\,\right]\,,
\label{eqapB1}
\end{eqnarray}
where we have used equation \eqref{eqtransv1}. Using the complex conjugate of equation \eqref{eqtransv1} in \eqref{eqapB1}, we finally obtain
\begin{equation}
\zeta_p^{-1}=\frac{\cos\phi_t}{\mu_0\mu_{2r}\omega}\cdot\left[\,u_t^{'}(1+\tan^2\phi_t)+iu_t^{''}(1-\tan^2\phi_t)\,\right]\,,
\label{eqapB2}
\end{equation}
and thus 
\begin{equation}
\text{Re}(\zeta_p^{-1})=\frac{u_t^{'}}{\mu_0\mu_{2r}\omega\cos\phi_t}=\frac{k_t^{'}\cos\theta_t}{\mu_0\mu_{2r}\omega\cos\phi_t}\,,
\label{eqapB3}
\end{equation}
which can be used in equation \eqref{EQP3} to show that its first term is zero.

The second identity is easier to show. Taking the real part of equation \eqref{eqzs0}, we obtain
\begin{equation}
\text{Re}(\zeta_s^{-1})=\frac{u_t^{'}}{\mu_0\mu_{2r}\omega\cos\phi_t}=\frac{k_t^{'}\cos\theta_t}{\mu_0\mu_{2r}\omega\cos\phi_t}\,,
\label{eqapB4}
\end{equation}
which shows that the second term of equation \eqref{EQP3} is also zero.

\bibliographystyle{epj}
\bibliography{EPJB}% Produces the bibliography via BibTeX.

\end{document}